\documentstyle[aasms4]{article}

\def\kms        {~km$\;$s$^{-1}$}

\begin{document}

\lefthead{Nonthermal Filaments as Cometary Wakes}
\righthead{SHORE \& LaROSA}

\title{The Galactic Center Isolated 
Nonthermal Filaments as Analogs of Cometary
Plasma Tails} 

\author{Steven N. Shore\altaffilmark{1} and T. N.
LaRosa\altaffilmark{2}{3}} 

\altaffiltext{1}{Department of Physics and Astronomy, Indiana
University South Bend, 1700 Mishawaka Avenue, South Bend, IN
46634-7111 (sshore$@$paladin.iusb.edu)} 

\altaffiltext{2}{Department of Biological and Physical Sciences,
Kennesaw State University, 1000 Chastain Road, Kennesaw, GA 30144
(ted$@$avatar.kennesaw.edu)} 

\altaffiltext{3}{NAvy-ASEE Summer Faculty Fellow, Naval Research Laboratory}

\begin{abstract} 

We propose a model for the origin of the isolated nonthermal filaments
observed at the Galactic center based on an analogy to cometary plasma
tails. We invoke the interaction between a large scale magnetized
galactic wind and embedded molecular clouds.  As the advected wind
magnetic field encounters a dense molecular cloud, it is impeded and
drapes around the cloud, ultimately forming a current sheet in the
wake. This draped field is further stretched by the wind flow into a
long, thin filament whose aspect ratio is determined by the balance
between the dynamical wind and amplified magnetic field pressures. 
The key feature of this cometary model is that the filaments are
dynamic configurations, not static structures.  The derived field
strengths for the wind and wake are consistent with observational
estimates.  Finally, the observed synchrotron emission is naturally
explained by the acceleration of electrons to high energy by plasma
and MHD turbulence generated in the cloud wake. 

\end{abstract} 


\newpage

\section{Introduction}

The isolated nonthermal filaments (hereafter, NTFs) in the Galactic
center (hereafter, GC) have remained unexplained since their discovery
by Morris and Yusef-Zadeh (1985).  It is generally accepted that these
are magnetic structures emitting synchrotron radiation since their
emission is strongly linearly polarized with the magnetic field
generally aligned with the long axis of the filaments (Bally \&
Yusef-Zadeh 1989; Gray et al. 1995; Yusef-Zadeh, Wardle, \& Parastaran
1997; Lang, Morris, \& Echevarria 1999). These structures are notable
for their exceptionally large length to width ratios, of order 10 to
100, and remarkable linearity (Yusef-Zadeh 1989; Morris 1996). 

To date, seven objects have been classified as NTFs. Six of these
point perpendicularly to the Galactic plane, but the most recently
discovered NTF is parallel to the plane (Anantharamaiah et al 1999).
The filaments have lengths up to 60 pc and often show feathering and
sub-filamentation on smaller transverse scale when observed at high
spatial resolution (Liszt and Spiker 1995; Yusef-Zadeh, Wardle, \&
Parastaran 1997; Lang, Morris \& Echevarria 1999; Anantharamaiah et al
1999).  The observed radio 20/90 cm spectral indices (defined as the
source flux, $S$, varying as $\nu^{\alpha}$) show a range, $-0.3 <
\alpha < -0.6$ (LaRosa et al 1999).  To date, there is no strong
evidence that the spectral index varies as a function of length along
the NTF (Lang, Morris \& Echevarria 1999; Kassim et al. 1999). Lastly,
it appears that all well studied NTFs may be associated with molecular 
clouds and/or H II regions (Serabyn \& Morris 1994; Uchida et al.
1996; Stahgun et al 1998). 

Several different types of models have been proposed for the
filaments.  These include magnetic field generation by an accretion
disk dynamo with subsequent transport of field into the interstellar
medium (Heyvaerts, Norman, \& Pudritz 1988); electrodynamic models of
molecular clouds moving with velocity ${\bf v}$ across a large scale
ordered magnetic field, ${\bf B}$, resulting in current formation by
${\bf v \times B}$ electric fields and subsequent pinching of these
currents into filaments (Benford 1988, 1997; Lesch \& Reich 1992);
magnetic reconnection between a molecular cloud field and the
large-scale ordered field (Serabyn \& Morris 1994); and particle
injection into interstellar magnetic field ropes at a stellar wind
termination shock (Rosner and Bodo 1996). Nicholls and Le Stange
(1995) proposed a specifically tailored model for G359.1-0.2, also
called the ``Snake''.  They invoke a high velocity star with a strong
stellar wind that is falling through the galactic disk, to create a
long wake, which they call a ``star trail''.  They must, however, fine
tune their model in order to obtain radio emission from the trail by
requiring the high energy electrons to be injected into the trail from
the supernova remnant G359.1-0.5. 

Although each of these models can in principle explain the particle
acceleration and radio emission, none except for the specialized star
trail scenario has satisfactorily accounted for the observed structure
of the filaments. For instance, Rosner and Bodo (1996) employ a
stellar wind termination shock as the source of high energy particles
that they assume are loaded onto pre-existing interstellar field
lines. The width of the resulting NTF is the radius of the stellar
wind bubble.  Synchrotron cooling leads, through a thermal
instability, to collapse of the filaments and amplification of the
internal magnetic field.  The streaming of the particles along these
otherwise quiescent field lines is assumed to produce the observed
long threads. However, MHD stability is a problem for this model, and
indeed for all models listed above, because magnetic fields left to
their own devices will deform through a rich variety of modes.  These
range from kink and sausage instabilities for ideal MHD to tearing
modes for resistive plasmas (e.g. Parker 1977; Cravens 1997). Unless
very special magnetic field configurations and boundary conditions are
imposed, and these are difficult to achieve even in the laboratory,
the length and thinness of the NTFs cannot be explained as static 
structures.  

In this paper, we adopt the viewpoint that the filaments are not
equilibrium structures but are rather dynamical structures embedded in
a flow. In a flow, the growth of many of the local instabilities is
suppressed by the advection. A similar conclusion has been reached by
Chandran, Cowley, \& Morris (1999) who argue that the filaments
represent locally illuminated regions of a large scale strong magnetic
field that is formed through the amplification of a weak {\it halo}
field by a galactic accretion flow.  In contrast, we propose an
alternative model in which the advection of a weak {\it galactic}
field in a large scale outflow from the central region is amplified
locally by encounters with interstellar clouds.   We find that several
key elements of the previously published scenarios are the natural
consequences of this cloud-wind interaction picture. 
 
\section{The Comet Model} 

The key feature of the physical interaction between a comet and the
magnetized solar wind was identified by Alfv\'en (1957) and elaborated
by many subsequent studies (e.g. Russell et al. 1991; Luhmann 1995;
Cravens 1997).  As the magnetic field that is carried in the wind
plasma encounters the comet, the field progress is retarded through
the coma because the magnetic diffusion times are much longer than the
advective timescale.   Mass loading of the solar wind from the coma
produces a velocity gradient, $\partial v_w/\partial x$, where $x$ is
the cross-tail direction.  The external field drapes over the coma and
is stretched by the wind, ultimately forming a current sheet in the
antisolar direction.  This field line draping, for a molecular cloud,
is depicted in Figure 1. 

Remote observations show that cometary streamers routinely display
aspect ratios of a 100 or more (Jockers 1991).  Direct {\it in situ}
plasma measurements of comet Giacobini-Zinner have confirmed the
overall picture of magnetic field draping.  In particular, these
encounter measurements show that the central tail axis consists of a
plasma sheet with very low magnetic field (Siscoe et al. 1986).  This
sheet is surrounded by a low density plasma that is threaded with the
draped wind magnetic field that has been compressed and amplified by
the flow.  Transverse pressure balance requires that the draped field
is about a factor of 5 to 10 stronger than the ambient field with
amplification occurring because of flux conservation and field line
stretching. This is our basic cometary analogy, that the ambient field
is anchored in the cloud and that the field line tilt and
amplification result from the shearing between the wake and the
external wind.  This picture, which explains solar system scale
phenomena rather well, is more than a mere analogy.  Any magnetized
wind that impacts a finite blunt body with low resistivity will
deflect around the object and drape the field along the wake flow. The
field diffusion time is $t_d = L^2/\eta$, where here $L$ is the cloud
radius and $\eta$ is the resistivity.  For the GC, this timescale is
many orders of magnitude larger than the wind advection time.
Consequently, the field evolution around the molecular cloud is
similar to the cometary case provided the field is ordered on the
cloud size scale, $L$.

We now explore the consequences of this scenario for the NTFs.
Consider a galactic scale wind with a mass loss rate $\dot{M}$.  This
wind need not eminate only from the GC.  With the broad spatial
distribution of star forming regions in the inner galaxy, we would
anticipate a roughly cylindrical -- not radial -- outflow which would
be sampled by clouds moving in whatever orbits they happen to have
relative to the plane.  Consequently, the wakes so produced should
generally be perpendicular to the plane.  For simplicity, however, we
will assume here a compact source.  The number density in the wind is
given by $n_W = \dot{M}v_{w,3}^{-1}r_{100}^{-2}$ cm$^{-3}$, for a mass
loss rate in $M_\odot$yr$^{-1}$, a wind speed $v_{w,3}$ in $10^3$\kms,
and a distance $r_{100}$ in 100 pc. For a cloud to survive in a
postulated galactic scale wind, its internal pressure must at least
balance the ram pressure of the background.  We assume that the cloud
pressure is given by $P= \rho_c\sigma_c^2$ where $\rho_c$ is the cloud
mass density and $\sigma_c$ its internal velocity dispersion. Hence,
for a wind of density $\rho_w$ and speed $v_w$, the required cloud
density is given by $\rho_c = \rho_w(\sigma/v_w)^2$. It has been 
inferred from {\it Ginga} and ASCA X-ray observations that the inner Galaxy
displays a strong wind (Yamauchi et al. 1990; Koyama et al. 1996). 
The average wind density within a radius of 80 pc of the GC is around
0.3 cm$^{-3}$ with a temperature of 10 keV and an expansion velocity
of about 3000 \kms (Koyama et al. 1996). These parameters correspond
to a mass loss rate of 10$^{-2}$ M$_\odot$ yr$^{-1}$ for a wind speed
of around 1000 km s$^{-1}$, which yields a critical cloud density of
order $n_c \ge 10^3$ cm$^{-3}$ for $\sigma = 20$ \kms (a typical
linewidth for molecular clouds in the GC region, see Morris \& Serabyn
[1996]) although clouds nearer the center will need higher densities
to survive.  This density estimate is a lower limit.  For clouds to
survive in the GC tidal field they must have densities at least an
order of magnitude above this (e.g. G\"usten 1989). The effect on the
cloud population is that massive, dense clouds will survive while
lower density, low mass clouds likely disperse on a dynamical
timescale and thus the cloud population may depend on galactocentric
distance. Dense clouds form wakes by geometrically blocking and
deflecting the wind.  We identify this wake, drawn out by the wind,
with the NTFs.  This scenario is sketched in Figure 1.  Thus follows
the essential predictive feature of our model: since the filaments are
not static structures, the classic MHD instabilities do not limit the
aspect ratio as they would for a static equilibrium field. 

What determines the structural properties of the wake, {\it i.e.} its
aspect ratio and length?  Given that $t_d \gg L/v_w$, the draped field
is stretched by the wind.  If $\delta v$ is the boundary layer shear
between the wind and cloud wake and $\Delta$ is a characteristic
length for the layer (of order $L$), then the axial field, $B_z$, as a
function of distance $Z$ behind the cloud is given by the induction
equation, $\partial B_z/\partial t = (\partial v/\partial x)B_0$,
where $B_0$ is the external field.  This has the approximate solution:
\begin{equation} 
B_z = B_0\frac{\delta v}{\Delta}\frac{z}{v_w}.
\end{equation} 
The axial field will continue to amplify until the draped magnetic
field pressure balances the ram pressure of the wind. In other words,
when the wake Alfv\'en speed equals the wind speed, $B_z/(4\pi
\rho_0)^{1/2} = v_w$, the field can no longer be stretched. This
provides a critical length: 
\begin{equation} z_c = 5\times
10^2n^{1/2}v_{w,3}^2\Delta /(B_{0,\mu}\delta v_3), 
\end{equation} 
where $v_{w,3} = v_w/10^3$km s$^{-1}$, $n$ is the number density, and
$B_{0,\mu}$ is the external field in $\mu$G. Thus, for $n \approx 1$
cm$^{-3}$ and $B_{0,\mu} \approx 10$, the predicted aspect ratio is
$z_c/\Delta \approx 50$.  Notice that stronger ambient fields lead to
shorter wakes. 

We now address the question of stability for the filaments. In the MHD
case, the velocity shear must exceed the Alfv\'en speed to produce a
growing mode for the Kelvin-Helmholtz instability (KHI).  Other
classical instabilities, such as the streaming, sausage, and kink
modes, have a similar criterion (e.g. Wang 1991).  Nonlinear models by
Malagoli, Bodo \& Rosner (1996) find that the fastest growing mode has
a wavenumber given by $k\Delta \approx 0.05$.  The KHI can therefore be
suppressed if the draped field amplification length is less than
$2\pi/k$.  Hence, for stability $z_c \le 40\pi \Delta$. With this
constraint, we find a lower limit on the external wind field strength,
$B_{0,\mu} \ge 40 n^{1/2}v_{w,3}/\delta v_3$.  Equipartition for the
wind plasma gives $B_{0,\mu} \approx 20$ for the parameters derived
from the ASCA data, which is in surprisingly good agreement with the
stability constraint.  Thus the expected amplified field strength is
$B_z \approx 2$ mG for $z/\Delta$ given by eq. (2).

Thus the key parameters that can be derived from our model are the
aspect ratio which depends on the wind parameters, and the magnetic
field strength in the filament.  The observed aspect ratios can be
explained using equation (2) with wind parameters consistent with the
ASCA data.  There are no direct measurments of the magnetic fields in
the filaments.  An estimate for the magnetic field can be derived from
the observed synchrotron luminosities using a minimum energy analysis.
The synchrotron luminosities are around  $10^{33} - 10^{34}$ erg
s$^{-1}$ (Gray et al. 1995; Lang, Morris, \& Echevarria 1999; Kassim et
al.  1999) and yield a magnetic field of $\sim 0.1$ mG about an order
of magnitude smaller than our model result.  Another estimate for the
field strength comes from assuming that the particles traverse the
length of an NTF in a time equal to their synchrotron lifetime.  The
synchrotron lifetime is $t_{\frac{1}{2}} = 1.20\times 10^4
B_{z,mG}^{-2}E_{GeV}^{-1}$ yrs, where $B_{z,mG}$ is the axial field in
milligauss and $E$ is the electron energy in GeV (e.g. Moffatt 1975).
Without reacceleration, assuming that the electrons are injected near
one end and radiate as they stream at the Alfv\'en speed (e.g. Wentzel
1974), the observed filament lengths give a field strength of 1 mG for
a length scale of 30 pc.  Fields strengths of 1 mG have also been
derived from dynamical arguments by Yusef-Zadeh and Morris (1987).  We
therefore conclude that our estimate of 1 mG is very reasonable and
that the minimum energy analysis of such structures which assumes
static and/or equilibrium conditions may produce misleading results.
Note that the synchrotron lifetime arguement indicates that
reacceleration or acceleration along the length of the filament is not
required, although as we now discuss acceleration along the filament is
expected in our picture.

Finally, since the NTFs are radiating via synchrotron emission, we
address the question of particle energization.  The observed emission
requires only a very small population of relativistic particles, of
order $10^{-5}$cm$^{-3}$.  The maximum energy that is available for
conversion to high energy particles is $VB_z^2/8\pi$, where $V$ is the
volume of the wake.  The maximum mean energy per particle that results
from this conversion is 10 GeV.  This is more than enough to explain
the radio emission.  A number of mechanisms that may be responsible
for particle acceleration are natural consequences of this MHD
configuration.  The wake must contain a current sheet.  Such
structures have been extensively studied in space plasmas.  The
simulation of sheared helmet plumes in the solar corona by Einaudi et
al. (1999) is particularly relevant to our scenario. They show that a
current sheet imbedded in a wake flow is unstable to the generation of
a local turbulent cascade without destruction of the large scale
advected structure.  Such cascades efficiently accelerate particles
through wave-particle interactions (Miller et al. 1997).  This
turbulent acceleration would therefore occur along the entire length
of the filament, thus spectral aging would not be observed in this
scenario. 

\section{Discussion and Conclusions} 

Santillan et al. (1999) have recently published numerical MHD
simulations of cloud collisions with a magnetized galactic disk.
Although these are ideal MHD and not of wind flow, they clearly
demonstrate that field line draping occurs as the interstellar clouds
move through a background large scale field.  In particular, their
Fig. 4 shows the formation of a narrow straight tail for the cloud
slamming into a transverse field imbedded in a planar gas layer.
Dynamically, this simulation differs from wind flow because the cloud
is slowed by the environmental gas.  Yet the essential physical
process is the same and closely resembles the simulations of cometary
tail evolution by Rauer et al. (1995). This cloud-wind interaction,
which may destroy the clouds if their masses are low enough (see
Vietri, Ferrara, \& Miniati 1997), is able to generate long magnetized
tails with large aspect ratios. 

Nonetheless, it has been argued in the literature that there is no
need for a dynamical explanation of these structures.   Recalling our
discussion of the various proposed static models for the NTFs, the
common explanation for their stability hinges on the existence of a
pervasive background field.  We see two ways of interpreting the
magnetic field measurements obtained from the filaments.  One is to
assume that they represent local enhacements of an otherwise weak, but
invisible, field.  The other is to assume that one is seeing a region
that happens to be locally illuminated but is otherwise extensive and
uniform.  We explicitly adopt the local enhancement picture and
propose a dynamical mechanism that can amplify the field to much
higher strength and still be stable.  On the other hand, assuming a
pervasive field still leaves the stability question unresolved for the
following reasons. A force-free equilibrium background field that is
presumably anchored in the turbulent gas of the galactic center
certainly will not be stable.  For instance, the solar corona has a
pervasive field that suffers both local and global instabilites.
Moreover, to stablize a filament, a pervasive field must have a
pressure gradient perpendicular to the filaments, so the field cannot
be uniform.  If it has gradients and curvature, a static magnetic
field is likely to be unstable.  In contrast, stability is not an
issue for a dynamical model, whether the flow is accreting (Chandran
et al. 1999) or, as in our case, an outflowing wind. 

The simplest geometry predicted by the cometary analogy is that every
filament should be associated with a molecular cloud on the side toward
the galactic plane.  This is seen for the Sgr C filament (Liszt \&
Spiker 1995) and the ``Snake'' (Uchida et al. 1996).  The model does
not, however, require this and more complex geometrical arrangements
are certainly possible in which environmental clouds interact with or
are superimposed on the filaments almost anywhere along their lengths.
 For instance, Yusef-Zadeh \& Morris (1987) find that a milligauss
field suffices to stabilize the filaments against ram pressure by
colliding molecular clouds.  We note that this is precisely the field
strength produced dynamically by the cometary model. 

In addition, a final state, where the cloud is completely dissipated,
could still permit the survival of the filament and has a cometary
analog. There are many instances in comets where the tail completely
separates from the coma and yet maintains structural coherence as it
is advected in the solar wind (e.g. Brandt \& Niedner 1987). These
so-called disconnection events could also occur in our picture.  In
such instances, there would be no cloud at either end of the filament.

We close by emphasizing that our aim here has been the exploration of
the consequences of a general scenario that can serve as a framework
for more quantitative calculations of the physical properties of the
Galactic Center filaments.  Although we use the special conditions at
the GC to constrain the mechanisms, the model is not constructed
specifically to explain the NTFs. Instead, they result from the
conditions that likely arise in any starburst galactic nucleus (see
Mezger, Duschl, \& Zykla 1996) and should be observable in such
environments.

\acknowledgments

We thank G. Einaudi, J. R. Jokipii, N. Kassim, C. Lang, J. Lazio, M.
Niedner, and M. Vietri for discussions, and  A. Santillan for
permission to quote his results prior to publication.  We especially
thank the referee, Mark Morris, for his critical reading of the
manuscript and for discussions, 
and B. D. G. Chandran for communicating his paper in advance of
publication. TNL is supported by a NAVY-ASEE faculty fellowship from
the Naval Research Laboratory and a NASA JOVE grant to Kennesaw State
University.  SNS is partially funded by NASA and thanks the
Astrophysics Group of the Physics Department of the University of Pisa
for a visiting appointment during summer 1998.

\newpage

Figure 1.  Schematic of the interaction of a magnetized wind
encountering a molecular cloud of radius L.  The wind velocity is
${\bf v_w}$ and the cloud velocity is ${\bf v_c}$.  The advected wind
magnetic field ${\bf B}_0$ is impeded by the cloud and we show how
successive field lines or flux ropes are strectched and draped by the
flow around the cloud into a long thin wake.  The draped field,
denoted ${\bf B}_z$, is oppositely directed in the wake and forms a
current sheet along the wake mid-plane.  Solar system studies indicate
that such a plasma-magnetic field configuration leads to particle
acceleration through turbulent dissipation and we therefore identify
such a wake as a nonthermal filament. 

\end{document}